\documentclass[aps,reprint,floatfix,nofootinbib,longbibliography,prx]{revtex4-2}
\usepackage[latin9]{inputenc}
\usepackage{verbatim}
\usepackage{mathrsfs}
\usepackage{amsmath}
\usepackage{amssymb}
\usepackage{amsfonts}
\usepackage{graphicx}



\begin{document}
\title{Helical structures in layered magnetic superconductors
due to  indirect exchange interactions mediated by interlayer tunneling}
\author{A. E. Koshelev}
\affiliation{Materials Science Division, Argonne National Laboratory, Lemont, Illinois
60439}
\begin{abstract}
Motivated by the recent discovery of helical magnetic structure in RbEuFe$_{4}$As$_{4}$, we investigate interlayer ordering of magnetic moments in materials composed of spatially-separated superconducting and ferromagnetically-aligned layers. We consider the interplay between the normal and superconducting indirect exchange interaction mediated by tunneling between the conducting layers. We elaborate a recipe to evaluate the normal interlayer interaction via two-dimensional density of states of an isolated layer and demonstrate that for bands with small fillings, such interaction is typically ferromagnetic and short-range. 
The nearest-layer interaction is proportional to the ratio of the interlayer hopping and in-plane band width squared.  
On the other hand, the superconducting contribution always gives antiferromagnetic interaction and may extend over several layers when the interlayer hopping energy exceeds the superconducting gap. The frustration caused by the interplay between the normal and superconducting parts may lead to spiral ground-state magnetic configuration. The four-fold in-plane anisotropy may lock the rotation angle between the moments in the neighboring layers to $90^\circ$, as it was observed in  RbEuFe$_{4}$As$_{4}$.
\end{abstract}
\date{\today }
\maketitle

\section{Introduction}

Europium-based iron pnictides have been introduced recently as a new platform to investigate the interplay between singlet superconductivity and magnetic order \citep{Ren2008,Jeevan2008,Jeevan2008a,RenPhysRevLett.102.137002,Jeevan2011,Cao2011,Liu2016a,Liu2016,KawashimaJPSJ2016,Zapf2017}.
Superconducting and ferromagnetic phases are two common electronic ground states
of conducting materials antagonistic to each other. Their mutual influence causes
many interesting phenomena which 
have been thoroughly investigated  
for a half century \citep{FuldeKellerInBook1982,BulaevskiiAdvPhys85,inAbrikosovBook,KulicBuzdinSupercondBook,MullerRoPP2001,GuptaAdvPhys06,WolowiecPhysC15}.

Coexistence of  uniform superconducting and ferromagnetic states is usually energetically unfavorable.  The coupling between these subsystems at the microscopic level is due to the exchange interaction of conducting electrons with localized moments. In normal state, such local coupling generates indirect interaction between the moments mediated by the mobile electrons known as Ruderman-Kittel-Kasuya-Yosida (RKKY) interaction \citep{RudermanPhysRev.96.99,*KasuyaPTP56,*YosidaPhysRev.106.893}.
Additionally, 
the superconducting response to the magnetic field generated by the aligned moments 
leads to the macroscopic electromagnetic interaction between the superconducting and magnetic subsystems 
\cite{KreyIJM72,*KreyIJM73,TachikiPhysRevB.20.1915}.  The common nonuniform ground-state configuration depends on the relative strength of these two interaction channels as well as on the relative strength of superconductivity and magnetism. 

The superconducting order strongly modifies the RKKY interaction at large distances. Anderson and Suhl  \citep{AndersonPhysRev.116.898} demonstrated that this modification may lead to ``cryptoferromagnetic'' state in which the ferromagnetic subsystem is split into a periodic array of small domains facilitating coexistence with superconductivity. The domain size is determined by the interplay between the short-range normal and long-range superconducting RKKY interactions. In isotropic materials, it is expected to be much smaller than the superconducting coherence length but much larger than the distance between the localized moments and the Fermi wavelength. 
A particular realization of such oscillatory magnetic state depends on the properties of the magnetic subsystem.
In the case of weak magnetic anisotropy, the helical 
structure may be formed instead. A detailed theory of such structure has been elaborated in Ref.~\citep{BulaevskiiJLTP1980} for the case of the isotropic electronic spectrum. %
Similar nonuniform magnetic states have also been predicted for the case of purely electromagnetic coupling between the subsystems  \citep{MatsumotoSSC79,GreensidePhysRevLett.46.49}.
Alternatively, a hard magnetic subsystem remaining uniform itself %
may make the superconducting system nonuniform via generation of spontaneous vortex lines \cite{KreyIJM72,*KreyIJM73,TachikiPhysRevB.20.1915}.

Several classes of superconducting materials with long-range magnetic order are known at present. In most materials, the magnetism is hosted by rare-earth elements which are sufficiently separated from the  conducting electrons so that the exchange interaction is relatively weak and does not destroy superconductivity.  Two first such groups of materials discovered in the 1970s are ternary molybdenum chalcogenides $\mathit{RE}$Mo$_6$\textit{X}$_8$ 
($\mathit{RE}$=rare-earth element and \textit{X}=S, Se) also known as Chevrel phases and ternary rhodium borides $\mathit{RE}$Rh$_4$B$_4$.  
The detailed description of their properties may be found in reviews  \citep{BulaevskiiAdvPhys85,WolowiecPhysC15} and in the book \citep{MapleFischerBook1982}. 
While most of these materials have antiferromagnetic order in the superconducting state, 
two notable exceptions are 
ErRh$_{4}$B$_{4}$ with $T_c\!\approx \! 8.5$K\citep{FertigPhysRevLett.38.987} and
Ho$_{1.2}$Mo$_6$S$_8$ with $T_c\!\approx \! 1.2$K\cite{IshikawaSSC77}, where superconductivity competes with the long-range ferromagnetic order. The emerging ferromagnetism at sub-Kelvin temperatures leads to the reentrance of the normal state  and formation of the intermediate oscillatory magnetic state in the narrow coexistence region, in qualitative agreement with theory \citep{AndersonPhysRev.116.898,BulaevskiiJLTP1980}. 
The similar reentrant behavior was also found in the rhodium stannide $\mathit{RE}$Rh$_{1.1}$Sn$_{3.6}$ \cite{RemeikaSSC1980}.

Two decades later, during the 1990s, the rare-earth nickel borocarbides $\mathit{RE}$Ni$_2$B$_2$C have been added to the family of magnetic  superconductors, see reviews \citep{MullerRoPP2001,GuptaAdvPhys06,MuellerHandbook2008,MazumdarPhysC2015}.
Distinctive features of these materials significantly expanded and enriched the field of coexistence of superconductivity and magnetic order. 
In contrast to the above cubic ternary compounds, these materials have layered structure: they are  composed of  conducting Ni layers and magnetic $\mathit{RE}$C layers.  In spite of such a structure, the electronic anisotropy is small. 
In compounds with $\mathit{RE}$=Tm, Er, Ho, and Dy, the superconductivity coexists with different kinds of magnetic order and every compound has some unique features. In most cases the magnetic moments are aligned within $\mathit{RE}$C layers and alternate from layer to layer (A-type antiferromagnets). In all compounds except the Tm one the moments are oriented along the layers. %
In HoNi$_2$B$_2$C the transition to such state occurs via the two intermediate incommensurate spiral configurations. %
Magnetic structure in ErNi$_2$B$_2$C is characterized by additional in-plane modulation, which is likely induced by coupling to the superconducting subsystem. 
In addition, a weak ferromagnetic state emerges in this material below 2.3K at the second magnetic transition, and this state coexists with superconductivity at lower temperatures.
%
%

All discussed materials are conventional singlet superconductors.
The triplet superconducting state, in which Cooper pairs are formed by electrons with the same spin, is less hostile to ferromagnetism than the singlet state.  The likely candidates for the triplet state are uranium-based compounds UGe$_2$, URhGe, and UCoGe discovered in the 2000s, see reviews \citep{AokiJPSJ12,*AokiJPSJ19,HuxleyPhysC2015}. The superconducting transitions in these compounds take place in sub-Kelvin temperature range, inside the ferromagnetic state. For such low transition temperatures, the superconductivity survives up to remarkably high magnetic field, 10-25 teslas, which is attributed to the triplet pairing. 

Recent development in the field is related to discovery and characterization of magnetically-ordered iron pnictides,
in particular, europium-based 122 compounds, see review \citep{Zapf2017}. 
As the borocarbides, these materials have layered structure: they are composed of the spatially separated magnetic Eu and conducting FeAs layers. The tunneling between the FeAs layers is rather strong so that the electronic anisotropy is also small. %
The parent non-superconducting material EuFe$_{2}$As$_{2}$
\citep{Ren2008,Jeevan2008a,Xiao2010} develops the spin-density wave order in the FeAs layers at 189K 
and the A-type antiferromagnetic order in the Eu$^{2+}$ layers at 19K with the magnetic moments oriented along the layers. It becomes superconducting under pressure with the maximum transition temperature reaching 30K at 2.6 GPa, so that the magnetic transition takes place in the superconducting state \citep{MicleaPhysRevB.79.212509,*TerashimaJPSJ2009}. %
Several substitutions also give superconductors coexisting with Eu magnetic order:
(i)K\citep{Jeevan2008} and Na\citep{QiNJP2008} on Eu site, (ii)Ir\cite{Paramanik2014}, Ru\citep{JiaoEPL2011}, and Co\cite{JiangPhysRevB.80.184514,*HeJPhys2010,*GuguchiaPhysRevB.84.144506,*JinPhysRevB.88.214516} on Fe site, and (iii)P on As site\citep{RenPhysRevLett.102.137002,Jeevan2011,Cao2011,TokiwaPhysRevB.86.220505,ZapfPhysRevLett.110.237002,NandiPhysRevB.89.014512}.
%
%
%
The maximum superconducting transition temperature for different substitution series ranges from 22 to 35K exceeding the magnetic-transition temperature in Eu layers. Therefore the unique feature of these materials is that they
exhibit Eu$^{2+}$ magnetism at the temperature scale, comparable with the superconducting transition.

The most investigated substituted compound is $\mathrm{Eu}\mathrm{Fe}_{2}(\mathrm{As}_{1-x}\mathrm{P}_{x})_{2}$. For optimal substitution, $x\approx 0.3$, it has the superconducting transition at 26K followed by the ferromagnetic transition at 19K. At lower temperatures, ferromagnetism coexists with superconductivity. 
In contrast to the parent compound, the Eu moments are oriented along the $c$ axis \citep{NandiPhysRevB.89.014512}. This leads to the formation of the composite domain and vortex-antivortex structure visualized by the decorations \citep{VeshchunovJETPLett2017} and magnetic-force microscopy \citep{StolyarovSciAdv18}.
This structure has been attributed to purely electromagnetic coupling between the magnetic moments and superconducting order parameter \citep{DevizorovaPhysRevLett.122.117002}. 
Alternatively, it also may be the realization of the ``cryptoferromagnetic'' state caused by the weak exchange interaction \citep{AndersonPhysRev.116.898}. 

The most recent development in the field is synthesis of the stoichiometric compounds \emph{A}EuFe$_{4}$As$_{4}$ with \emph{A}=Rb \citep{Liu2016a,KawashimaJPSJ2016,Bao2018,Smylie2018,Stolyarov2018} and Cs \citep{Liu2016,KawashimaJPSJ2016} in which every second layer of Eu is completely substituted with Rb or Cs. These 1144 materials have the superconducting transition temperature of 36.5 K, higher than the doped 122 Eu compounds. On the other hand, the magnetic transition temperature is 15K, which is 4K lower than in EuFe$_{2}$As$_{2}$,  most likely because of the weaker interaction between the magnetic layers. These materials are characterized by low superconducting anisotropy $\sim 1.7$ and highly-anisotropic easy-axis Eu magnetism  \citep{Smylie2018,WillaPhysRevB.99.180502}. 
With increasing pressure the superconducting temperature is suppressed and the magnetic temperature is enhanced so that they cross at $~\sim 7$GPa and at higher pressures superconducting transition takes place in the magnetically-ordered state \cite{JacksonPhysRevB.98.014518,XiangPhysRevB.99.144509}. 
Recent resonant X-ray scattering and neutron diffraction measurements revealed that the magnetic structure is 
helical: the Eu moments rotate  90$^\circ$ from layer to layer \citep{IidaPhysRevB.100.014506,IslamPreprint2019}, see the picture in Fig.~\ref{Fig:XSCn}.

Motivated by this unexpected finding, we investigate magnetic structure in a material composed of spatially separated ferromagnetically-aligned and superconducting layers.  
Due to the large separation, the interaction between different Eu layers in \emph{A}EuFe$_{4}$As$_{4}$ 
most likely has the RKKY origin. 
As the Eu 4f orbitals are strongly localized, one can only consider interaction with the closest FeAs layers.  An essential feature of the \emph{A}EuFe$_{4}$As$_{4}$ structure is that the neighboring Eu layers have a direct coupling with different FeAs layers, see the picture in Fig.~\ref{Fig:XSCn}. Consequently, the magnetic interlayer interaction is mediated by tunneling between the conducting layers.  As these materials are not very anisotropic, this mechanism is not particularly weak. 
In contrast, the Eu-122 compounds and borocarbides are composed of alternating magnetic and conducting layers. In this case, two adjacent magnetic layers couple with the same conducting layer yielding a finite interlayer interaction even without tunneling between the conducting layers. 

One may think that a possible alternative to the RKKY mechanism may be the electromagnetic dipole interactions.
We note, however, that the dipole interaction between uniformly polarized layers 
is very small when separation between them exceeds the in-plane distance between the moments $a$. Indeed, the average magnetic field outside of a uniformly polarized layer is zero and the oscillating component decays away from the layer very fast, as $\exp(-2\pi z/a)$ \cite{BensonPhysRev.178.839,*MaleevJETP76}. For distances $z$ much smaller than the London penetration depth, superconductivity has a very weak influence on this behavior.  In the case of \emph{A}EuFe$_{4}$As$_{4}$, $a=3.9 \AA$ and the separation between the magnetic layers is $c=13.3\AA$ \cite{Liu2016a}. This means that the exponential factor is estimated as  $\exp(-2\pi c/a)\approx 4.6 \cdot 10^{-10}$, i.e., the dipole interaction is negligible even for neighboring layers. 

%
We consider magnetic structure appearing as a result of the interplay between the normal and superconducting RKKY interactions mediated by tunneling between the conducting layers. 
To highlight essential physics, we mostly study a relatively simple single-band model with open Fermi surface. 
We relate the normal interlayer interaction with the two-dimensional density of states of an isolated layer. 
An important observation is that this interaction vanishes for quadratic in-plane spectrum corresponding to the energy-independent density of states. In the case of a shallow band when non-parabolicity is small, the normal-state interlayer interaction is ferromagnetic and short-range. %
Such behavior is very different from the oscillatory in-plane RKKY interaction, which has been investigated for the iron pnictides and selenides in Ref.~\citep{AkbariNJP2013}. %
The superconducting contribution always gives antiferromagnetic interlayer interaction that may extend over several layers when the interlayer hopping exceeds the superconducting gap. As a result of frustrating interlayer interactions caused by the interplay between the normal and superconducting parts, the ground-state magnetic configuration may be a helix.  
A similar physical mechanism of the helical magnetic structure in
RbEuFe$_{4}$As$_{4}$ has been proposed in the recent paper \cite{DevizorovaPhysRevB.100.104523}
based on the earlier results obtained for isotropic electronic spectrum
\citep{BulaevskiiJLTP1980}. We point, however, that the layered structure
and open Fermi surface lead to qualitative modifications of both normal
and superconducting RKKY interactions, which are in the focus of this
paper. %
In general, the rotation angle between the moments in the neighboring layers continuously varies with the model parameters. The 90$^\circ$ helix observed in RbEuFe$_{4}$As$_{4}$ is most likely related to the in-plane four-fold anisotropy which locks such structure within a finite range of parameters.

The paper is organized as follows: In Sec.~\ref{sec:Model}, we introduce the model. In Secs.~\ref{sec:Normal}  and \ref{sec:Supercond}, we consider the normal and superconducting interactions between magnetic layers mediated by the indirect  exchange due to tunneling between the conducting layers.
In Sec.~\ref{sec:multiband}, we discuss generalization of these results to multiple-band materials. 
In Secs.~\ref{sec:HelixEn} and \ref{sec:OptMod}, we compute the energy of helical structure and the optimal modulation vector.


\section{Model}\label{sec:Model}

We consider a material composed of superconducting and magnetic layers. The local moments inside the magnetic layers are assumed to be ordered ferromagnetically. The major focus of this paper is the interlayer magnetic order emerging due to the interplay between the normal and superconducting RKKY interactions. The system under consideration is described by the Hamiltonian
\begin{equation}
\hat{H}=\hat{H}_{S}+\hat{H}_{M}+\hat{H}_{MS},\label{eq:Hamilt}
\end{equation}
where the first term
\begin{align}
\hat{H}_{S} & =\sum_{n}\int d^{2}\boldsymbol{r}\left[\psi_{n,\alpha}^{\dagger}(\boldsymbol{r})\hat{\xi}_{2D}\psi_{n,\alpha}(\boldsymbol{r})\right.\nonumber \\
- & t_{\bot}\left(\psi_{n,\alpha}^{\dagger}(\boldsymbol{r})\psi_{n\!+\!1,\alpha}(\boldsymbol{r})+\psi_{n\!+\!1,\alpha}^{\dagger}(\boldsymbol{r})\psi_{n,\alpha}(\boldsymbol{r})\right)\nonumber \\
 & -\left.\frac{g}{2}\psi_{n,\alpha}^{\dagger}(\boldsymbol{r})\psi_{n,\beta}^{\dagger}(\boldsymbol{r})\psi_{n,\beta}(\boldsymbol{r})\psi_{n,\alpha}(\boldsymbol{r})\right]\label{eq:HS}
\end{align}
describes the superconducting subsystem. Here $\alpha,\beta$ are spin indices,  $\hat{\xi}_{2D}=\xi_{2D}\left(\hat{\boldsymbol{p}}\right)=\varepsilon_{2D}\left(\hat{\boldsymbol{p}}\right)-\varepsilon_{F}$
is the single-layer spectrum, and $t_{\bot}$ is the interlayer hopping energy. For bands with low fillings, this
spectrum can be approximated as quadratic, $\varepsilon_{2D}\left(\boldsymbol{p}\right)=p^{2}/2m$.
The full three-dimensional electronic spectrum of this model is 
\begin{equation}
\varepsilon(\boldsymbol{p},q)=\varepsilon_{2D}\left(\boldsymbol{p}\right)+\varepsilon_{z}(q),
\label{eq:3Dspectrum}
\end{equation}
where for the assumed simplest interlayer tunneling $\varepsilon_{z}(q)=-2t_{\bot}\cos q$
with $q$ being the reduced c-axis wave vector, $-\pi\!<\!q\!<\!\pi$. 
Such spectrum for $\varepsilon_{F}\!>\!2t_{\bot}$ corresponds to open Fermi surface.  
The second term $\hat{H}_{M}$ in Eq.\ \eqref{eq:Hamilt} describes the magnetic layers favoring ferromagnetic alignment of the moments inside them. 
Its particular form does not play a role in our consideration. 
In the case of RbEuFe$_{4}$As$_{4}$, the thermodynamics of the magnetic transition is well described by the easy-plane quasi-two-dimensional Heisenberg model \citep{WillaPhysRevB.99.180502}.

The interaction between the superconducting and magnetic subsystems is determined by the local exchange Hamiltonian
\begin{equation}
\hat{H}_{MS} \!=\!\sum_{n,m}\!\int\!\! d^{2}\!\boldsymbol{r}
\!\sum_{\boldsymbol{R}}
\psi_{n,\alpha}^{\dagger}(\boldsymbol{r})J_{nm}(\boldsymbol{r}\!-\!\boldsymbol{R})\boldsymbol{\sigma}_{\alpha\beta}\boldsymbol{S}_{m}(\boldsymbol{R})\psi_{n,\beta}(\boldsymbol{r}),\label{eq:HMS}
\end{equation}
where $\boldsymbol{S}_{m}(\boldsymbol{R})$ are localized spins and $\boldsymbol{\sigma}_{\alpha\beta}$ is the Pauli-matrix vector. We can rewrite $\hat{H}_{MS}$ as 
\[
\hat{H}_{MS}=-\sum_{n}\int d^{2}\boldsymbol{r}\psi_{n,\alpha}^{\dagger}(\boldsymbol{r})\boldsymbol{h}_{n}(\boldsymbol{r})\boldsymbol{\sigma}_{\alpha\beta}\psi_{n,\beta}(\boldsymbol{r}),
\]
where 
\begin{equation}
\boldsymbol{h}_{n}(\boldsymbol{r})=-\sum_{m}\sum_{\boldsymbol{R}}J_{nm}(\boldsymbol{r}-\boldsymbol{R})\boldsymbol{S}_{m}(\boldsymbol{R})\label{eq:ExField}
\end{equation}
is the effective exchange field acting on spins of conducting electrons\footnote{The effective exchange field caused by nonuniform magnetic field $\boldsymbol{H}(\boldsymbol{r})$ is $\boldsymbol{h}(\boldsymbol{r})=\mu_0\boldsymbol{H}(\boldsymbol{r})$, where $\mu_0$ is the electron's magnetic moment}.
The fermionic response to such a field in both normal and superconducting
state is determined by the nonlocal spin susceptibility $\chi_{n}(\boldsymbol{r})$
and at fixed arrangements of the localized moments the corresponding
energy contribution is 
\begin{equation}
E_{MS}=\!-\frac{1}{2}\sum_{n,n^{\prime}}\!\int \!\!d^{2}\boldsymbol{r}\int \!\!d^{2}\boldsymbol{r}^{\prime}
\chi_{n-n^{\prime}}(\boldsymbol{r}\!-\!\boldsymbol{r}^{\prime})\boldsymbol{h}_{n}(\boldsymbol{r})\boldsymbol{h}_{n^{\prime}}(\boldsymbol{r}^{\prime}).\label{eq:EMS}
\end{equation}
In superconducting state, the assumed linear-response approximation is valid if the amplitude of the exchange field $h$ is smaller than the superconducting gap $\Delta$. 

In this paper we focus on the case when the localized spins are ferromagnetically
aligned inside the layers, i.e., $\boldsymbol{S}_{m}(\boldsymbol{R})$
is in-plane coordinate independent. In this case $\boldsymbol{h}_{n}(\boldsymbol{r})$
also becomes uniform, $\boldsymbol{h}_{n}(\boldsymbol{r})\!\rightarrow\!\boldsymbol{h}_{n}\!=\!-\sum_{m}\mathcal{J}_{nm}\boldsymbol{S}_{m}$
with $\mathcal{J}_{nm}\!=\!\sum_{\boldsymbol{R}}J_{nm}(\boldsymbol{r}\!-\!\boldsymbol{R})$
being the total exchange interaction from all aligned spins in a single
layer. In this case, Eq.~\eqref{eq:EMS} gives the following result
for the energy per unit area per magnetic layer
\begin{align}
F_{MS}=&-\frac{1}{2N_{M}}\sum_{n,n^{\prime}}\mathcal{X}_{n-n^{\prime}}\boldsymbol{h}_{n}\boldsymbol{h}_{n^{\prime}}\nonumber\\
=&-\frac{1}{2N_{M}}\sum_{m,m^{\prime}}\mathcal{I}_{m-m^{\prime}}\boldsymbol{S}_{m}\boldsymbol{S}_{m^{\prime}}
\label{eq:FMS}
\end{align}
where $N_{M}$ is the total number of magnetic layers, 
\begin{equation}
\mathcal{X}_{n}=\int d\boldsymbol{r}\chi_{n}(\boldsymbol{r})\label{eq:IntSuscDef}
\end{equation}
is the nonlocal susceptibility integrated over the in-plane coordinate,
and 
\begin{equation}
\mathcal{I}_{m-m^{\prime}}=\sum_{n,n^{\prime}}\mathcal{X}_{n-n^{\prime}}\mathcal{J}_{nm}\mathcal{J}_{n^{\prime}m^{\prime}}\label{eq:EffExchange}
\end{equation}
are the effective interlayer interaction constants. Being mediated by the conduction electrons, these constants have an RKKY nature. For brevity, in the following consideration we refer to the quantity $\mathcal{X}_{n}$ in Eq.~\eqref{eq:IntSuscDef} as an interlayer susceptibility. 

We consider the situation when (i)RKKY is the dominating mechanism and therefore Eqs.~\eqref{eq:FMS} and \eqref{eq:EffExchange} determine the interlayer magnetic structure and (ii)different magnetic layers are not coupled with the same metallic layer, i.e., $\sum_{n}\mathcal{J}_{nm}\mathcal{J}_{nm^{\prime}}=0$ for $m\neq m^{\prime}$ and the interlayer interactions only appear due to tunneling between the metallic layers yielding finite $\mathcal{X}_{n}$ for $n\neq0$. The latter situation is realized in the magnetic 1144 iron arsenides. In the case of RbEuFe$_{4}$As$_{4}$, due to the strongly localized nature of the $4f$ states, one can only take into account the exchange interactions with closest metallic layers. Therefore, we may only consider interactions of a Eu layer with index $m$ with the two neighboring FeAs layers with the indices
$n=2m-1$ and $2m$, see the picture in Fig.~\ref{Fig:XSCn}. Dropping the indices in these nearest-neighbor exchange constants $\mathcal{J}_{nm}$, we evaluate from Eq.~\eqref{eq:EffExchange} 
\begin{equation}
\mathcal{I}_{l}\!\approx\!\mathcal{J}^{2}\!\!\sum_{\delta,\delta^{\prime}=0,1}\!\!\mathcal{X}_{2l-\delta+\delta^{\prime}}
\!=\!\mathcal{J}^{2}\left(\mathcal{X}_{2l\!-\!1}\!+\!2\mathcal{X}_{2l}\!+\!\mathcal{X}_{2l\!+\!1} \right).
\label{eq:IlXl1144}
\end{equation}
Thus the interlayer interaction is determined by the interlayer susceptibility. 
We proceed with the evaluation of the normal and superconducting contributions to this key quantity.

\section{Normal-state interlayer susceptibility}\label{sec:Normal}

The indirect interaction between localized magnetic moments in metals mediated by conducting electrons is known as the RKKY interaction \citep{RudermanPhysRev.96.99,*KasuyaPTP56,*YosidaPhysRev.106.893}. Its shape is determined by the nonlocal spin susceptibility, which in the standard case of closed Fermi surface has a well-known oscillating behavior $\propto-\cos\left(2p_{F,r}r\right)/r^{3}$, where $p_{F,r}$ is the Fermi momentum along the considered direction\citep{RothPhysRev.149.519}. The behavior for a layered material with the spectrum in Eq.~\eqref{eq:3Dspectrum} corresponding to open Fermi surface is qualitatively different. In this case the spin susceptibility $\chi_{n}(\boldsymbol{r})$ in Eqs.~\eqref{eq:EMS} and \eqref{eq:IntSuscDef} is
\begin{align}
\chi_{n}(\boldsymbol{r}) & \!=\!-\!2\int\frac{d^{2}\boldsymbol{p}dq}{(2\pi)^{3}}\int\frac{d^{3}\boldsymbol{p}^{\prime}dq^{\prime}}{(2\pi)^{3}}\frac{f(\boldsymbol{p},q)\!-\!f(\boldsymbol{p}^{\prime},q^{\prime})}{\varepsilon(\boldsymbol{p},q)\!-\!\varepsilon(\boldsymbol{p}^{\prime},q^{\prime})}\nonumber \\
\times & \exp\left[i\left(\boldsymbol{p}-\boldsymbol{p}^{\prime}\right)\boldsymbol{r}+i\left(q-q^{\prime}\right)n\right],\label{eq:SuscLay}
\end{align}
where $f(\boldsymbol{p},q)=\left\{ 1+\exp\left[\left(\varepsilon(\boldsymbol{p},q)-\varepsilon_{F}\right)/T\right]\right\} ^{-1}$
is the Fermi-Dirac distribution function and the energy $\varepsilon(\boldsymbol{p},q)$ is given by Eq.~\eqref{eq:3Dspectrum}. 
The nonlocal susceptibility of layered metal has been evaluated in Ref.~\citep{AristovPhysRevB.55.8064}. It oscillates both as a function of  $\boldsymbol{r}$ at fixed $n$ and as a function of layer index $n$ at fixed $\boldsymbol{r}$. %
Here we reconsider this problem with the goal to directly evaluate the interlayer
susceptibility in Eq.~\eqref{eq:IntSuscDef}. Using the above formula
for $\chi_{n}(\boldsymbol{r})$, we derive the convenient representation
\begin{align}
\mathcal{X}_{n}^{N}\! & =\!-\!\int\limits _{-\pi}^{\pi}\frac{dq}{2\pi}\int\limits _{-\pi}^{\pi}\frac{dq^{\prime}}{2\pi}\int d\xi\nu_{2D}(\xi)\nonumber \\
\times & \frac{f(\xi\!+\!\varepsilon_{z})-f(\xi\!+\!\varepsilon_{z}^{\prime})}{\varepsilon_{z}-\varepsilon_{z}^{\prime}}\exp\left[i\left(q-q^{\prime}\right)n\right],\label{eq:IntSuscGen}
\end{align}
where 
\begin{equation}
\nu_{2D}(\xi)=2\int\frac{d^{2}\boldsymbol{p}}{(2\pi)^{2}}\delta\left(\xi\!-\!\varepsilon_{2D}\left(\boldsymbol{p}\right)\!+\!\varepsilon_{F}\right)\label{eq:2DDoSDef}
\end{equation}
is the total density of states for an isolated layer for both spin orientations, $f(\xi)\!=\!\left[1\!+\!\exp\left(\xi/T\right)\right]^{-1}$,
and we use the abbreviations $\varepsilon_{z}\!=\!\varepsilon_{z}(q)$ and
$\varepsilon_{z}^{\prime}\!=\!\varepsilon_{z}(q^{\prime})$. We will be
mostly interested in the zero-temperature limit for which we obtain
\begin{align}
\mathcal{X}_{n}^{N} & \!=\!-\int\limits _{-\pi}^{\pi}\!\frac{dq}{2\pi}\int\limits _{-\pi}^{\pi}\!\frac{dq^{\prime}}{2\pi}\!\int\limits _{-\varepsilon_{z}^{\prime}}^{-\varepsilon_{z}}\!d\xi\frac{\nu_{2D}(\xi)}{\varepsilon_{z}\!-\!\varepsilon_{z}^{\prime}}\exp\left[i\left(q\!-\!q^{\prime}\right)n\right].\label{eq:IntSuscEnDep}
\end{align}
Note that the sum $\sum_{n=-\infty}^{\infty}\mathcal{X}_{n}^{N}$ gives
the uniform susceptibility equal to $\nu_{2D}(0)$. 

For quadratic in-layer spectrum, we have the well-known energy-independent 2D DoS $\nu_{2D}(\varepsilon,q)=m/\pi$. In this case the above equation immediately gives $\mathcal{X}_{n}^{N}=\nu_{2D}\delta_{n}$, i.e., \emph{for the quadratic spectrum the RKKY interaction between ferromagnetically-ordered layers is absent}. The oscillating coordinate dependence of $\chi_{n}(\boldsymbol{r})$ has been evaluated in Ref.~\citep{AristovPhysRevB.55.8064}. It is crucial that the integral of this function over $\boldsymbol{r}$ vanishes for $n\neq0$ and this property has important implications for the RKKY interaction between ferromagnetically-ordered layers. 

In general, the density of state is energy dependent. When variation
of $\nu_{2D}(\xi)$ at the scale $\xi\sim\varepsilon_{z}$ are weak,
we can use expansion with respect to derivatives of $\nu_{2D}(\xi)$
near the Fermi level 
\[
\int\limits _{-\varepsilon_{z}^{\prime}}^{-\varepsilon_{z}}\!d\xi\nu_{2D}(\xi)\!=\!\sum_{s=0}^{\infty}\nu_{2D}^{(s)}\frac{(-\varepsilon_{z})^{s+1}\!-(-\varepsilon_{z}^{\prime})^{s+1}}{\left(s+1\right)!},
\]
with $\nu_{2D}^{(s)}\equiv d^{s}\nu_{2D}/d\xi^{s}$ at $\xi=0$, which yields
\begin{align*}
\mathcal{X}_{n}^{N} & \!=\!\sum_{s=0}^{\infty}\frac{\left(-1\right)^{s}\nu_{2D}^{(s)}}{\left(s+1\right)!}\\
\times & \int\limits _{-\pi}^{\pi}\!\frac{dq}{2\pi}\int\limits _{-\pi}^{\pi}\!\frac{dq^{\prime}}{2\pi}\frac{\varepsilon_{z}^{s+1}\!-\!\varepsilon_{z}^{\prime s+1}}{\varepsilon_{z}\!-\!\varepsilon_{z}^{\prime}}\exp\left[i\left(q\!-\!q^{\prime}\right)n\right].
\end{align*}
The second-derivative term ($s\!=\!2$) in this sum, \[\frac{\nu_{2D}^{\prime\prime}}{6}\left(2\delta_{n}\int_{-\pi}^{\pi}\frac{dq}{2\pi}\varepsilon_{z}^{2}+\left|\int_{-\pi}^{\pi}\frac{dq}{2\pi}\varepsilon_{z}\exp\left(iqn\right)\right|^{2}\right),\]
gives a finite nearest-neighbor interlayer interaction. For
$\varepsilon_{z}=-2t_{\bot}\cos q$, we have $\int_{-\pi}^{\pi}\frac{dq}{2\pi}\varepsilon_{z}\exp\left(iqn\right)=-t_{\bot}\delta_{|n|-1}$
and 
\begin{equation}
\mathcal{X}_{1}^{N}\approx\frac{\nu_{2D}^{\prime\prime}}{6}t_{\bot}^{2}.\label{eq:NN-SecondDir}
\end{equation}
The interaction is ferromagnetic if $\nu_{2D}^{\prime\prime}>0$.
As $\nu_{2D}^{\prime\prime}\approx\nu_{2D}/W^{2}$, where $W$ is
the in-plane band width, $\mathcal{X}_{1}^{N}\sim\mathcal{X}_{0}^{N}t_{\bot}^{2}/W^{2}$$\ll\mathcal{X}_{0}$.
For larger $n$, $\mathcal{X}_{n}^{N}$ steeply decreases as $\mathcal{X}_{0}^{N}\left(t_{\bot}^{2}/W^{2}\right)^{n}$meaning
that the terms with $n>1$ can be safely neglected. Such behavior
is very different from the conventional oscillating behavior with
the power envelope decrease, which is realized for closed Fermi surfaces. 

For example, for the quartic correction to the spectrum in the form
\begin{equation}
\varepsilon_{\boldsymbol{p}}^{(2D)}=\frac{p^{2}}{2m}+\frac{\alpha\left(p_{x}^{4}+p_{y}^{4}\right)}{4m^{2}}\label{eq:QuarticSpectrum}
\end{equation}
the density of states is
\begin{align*}
\nu_{2D}(\xi) & =\frac{m}{\pi}\int_{0}^{\pi}\frac{d\theta}{\pi}\frac{1}{\sqrt{1+\alpha(\varepsilon_{F}+\xi)\left(3+\cos\theta\right)}}.
\end{align*}
The expansion for small $\xi$ valid for $\alpha\varepsilon_{F}\ll1$
gives
\begin{equation}
\nu_{2D}^{\prime\prime}\approx\frac{m}{\pi}\frac{57}{8}\alpha^{2}.
\end{equation}
In this case $\nu_{2D}^{\prime\prime}>0$ meaning that the spectrum
in Eq.~\eqref{eq:QuarticSpectrum} favors the ferromagnetic alignment
between the layers, independently of sign of the quartic coefficient $\alpha$.

\section{Superconducting contribution to interlayer susceptibility}\label{sec:Supercond}

\begin{figure}
	\includegraphics[width=3.4in]{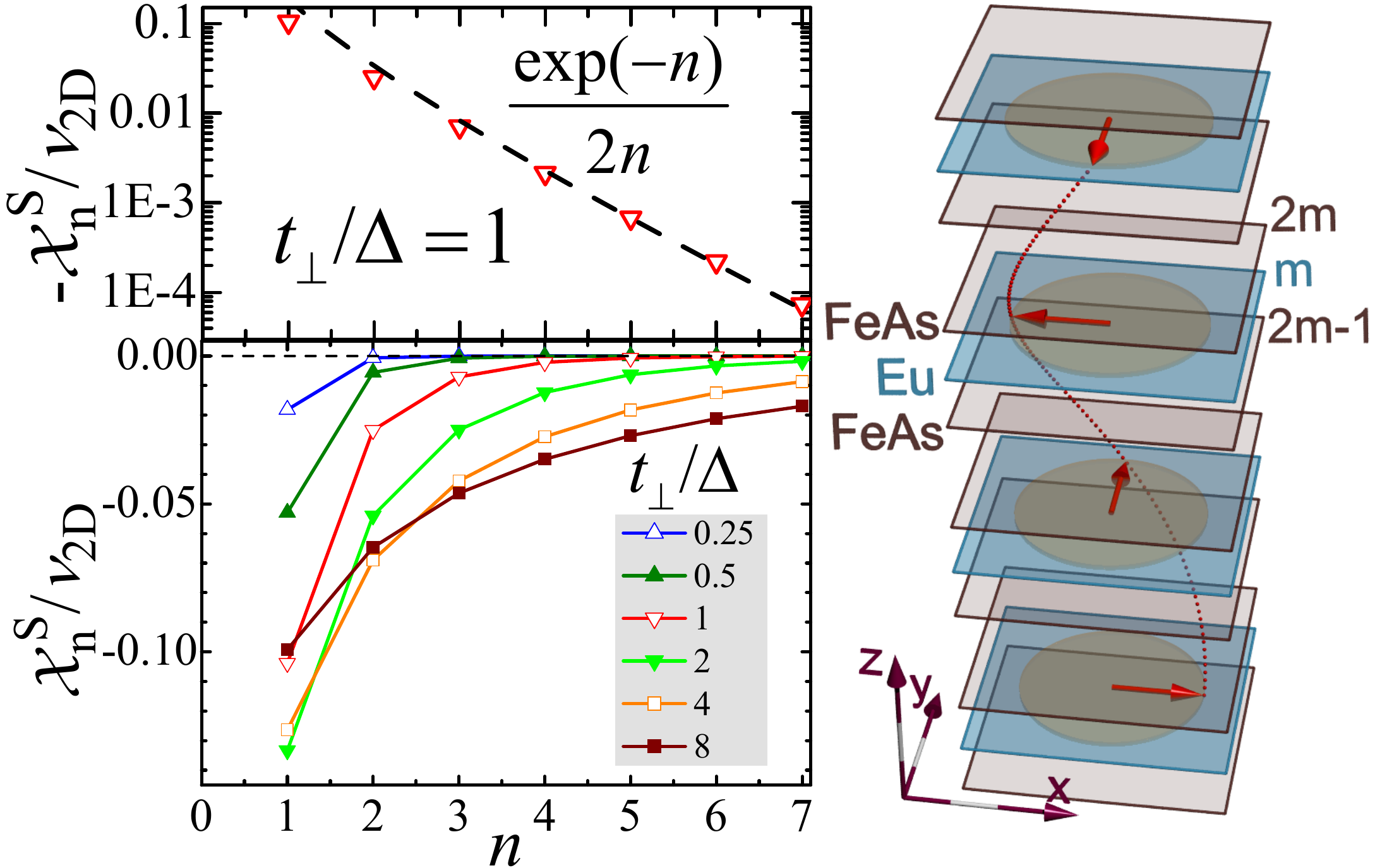}
	\caption{Lower left: The plots of the superconducting contribution to the interlayer susceptibility
		$\mathcal{X}_{n}^{S}$ normalized to DoS $\nu_{2D}$, Eq.~\eqref{eq:IntSuscSCPres},
		for different ratios $t_{\bot}/\Delta$. The upper left figure shows the semilog
		plot of $-\mathcal{X}_{n}^{S}/\nu_{2D}$ for $t_{\bot}/\Delta=1$ together with the analytical large-$n$
		asymptotics. The picture on the right illustrates the arrangement of the magnetic Eu and conducting FeAs layers in RbEuFe$_{4}$As$_{4}$ and the interlayer helical magnetic order.
	}
	\label{Fig:XSCn}
\end{figure}
We proceed with evaluation of the superconducting contribution to the interlayer susceptibility.   A general formula for the nonlocal spin susceptibility in the superconducting state is
\begin{align}
\chi(\boldsymbol{r},n) & \!=\!-2T\sum_{\omega_{s}}\left[\mathscr{\mathcal{G}}^{2}\left(\omega_{s},\boldsymbol{r},n\right)\!+\!\left|\mathcal{F}\left(\omega_{s},\boldsymbol{r},n\right)\right|^{2}\right],\label{eq:SuscSCDef}
\end{align}
see, e.g., Ref.~\cite{KochelaevJETP79}.
Here $\omega_s=\pi T(2s+1)$ are the Matsubara frequencies, 
\begin{align*}
\mathcal{G}\left(\omega_{s},\boldsymbol{r},n\right) & \!=\!\int_{\boldsymbol{p},q}\!\!
\exp\left(i\boldsymbol{p}\boldsymbol{r}\!+\!iqn\right)\frac{i\omega_{s}+\xi(\boldsymbol{p},q)}
{\omega_{s}^{2}\!+\!\left[\xi(\boldsymbol{p},q)\right]^{2}\!+\!\Delta^{2}},\\
\mathcal{F}\left(\omega_{s},\boldsymbol{r},n\right) & \!=\!\int_{\boldsymbol{p},q}\!\!
\exp\left(i\boldsymbol{p}\boldsymbol{r}\!+\!iqn\right)
\frac{\Delta}{\omega_{s}^{2}\!+\!\left[\xi(\boldsymbol{p},q)\right]^{2}\!+\!\Delta^{2}}
\end{align*}
are the regular and anomalous Green's functions with $\int_{\boldsymbol{p},q}\equiv \int\!\frac{d^{2}\boldsymbol{p}}{(2\pi)^{2}}\int_{-\pi}^{\pi}\!\frac{dq}{2\pi}$, and $\Delta$ is the superconducting gap.
Therefore, the interlayer susceptibility, Eq.~\eqref{eq:IntSuscDef}, is given by
\begin{equation}
\mathcal{X}_{n}  \!=\!-2T\sum_{\omega_{s}}\!\int \!d\boldsymbol{r}
\left[\mathcal{G}^{2}\left(\omega_{s},\boldsymbol{r},n\right)\!+\!\left|\mathcal{F}\left(\omega_{s},\boldsymbol{r},n\right)\right|^{2}\right].\label{eq:IntSuscSCDef}
\end{equation}
Note that the linear response with respect to the effective
field $h$ assumed in derivation of Eqs.~\eqref{eq:EMS}--\eqref{eq:EffExchange} is valid for $h\ll\Delta$. 
We consider again the case of open Fermi surface described by the spectrum $\xi(\boldsymbol{p},q)=p^{2}/2m+\varepsilon_{z}(q)-\varepsilon_{F}$.
The derivations described in
Appendix \ref{App:DerivSCSusc} lead to the following presentation
for the superconducting contribution to $\mathcal{X}_{n}$
\begin{align}
\mathcal{X}_{n}^{S} & =\!\nu_{2D}\,\mathcal{S}_{n}(2t_{\bot}/\Delta), \label{eq:IntSuscSCPres} \\
\mathcal{S}_{n}(\tau) & =-\int\limits _{0}^{\pi}\frac{dq_{+}}{\pi}\int\limits _{0}^{\pi}\frac{dq_{-}}{\pi}\cos\left(q_{-}n\right)\nonumber\\
\times & \frac{\ln\left(\sqrt{1\!+\!\tau^{2}\sin^{2}q_{+}\sin^{2}\left(\frac{q_{-}}{2}\right)}\!+\!\tau\sin q_{+}\sin\left(\frac{q_{-}}{2}\right)\right)}{\tau\sin q_{+}\sin\left(\frac{q_{-}}{2}\right)\sqrt{1+\tau^{2}\sin^{2}q_{+}\sin^{2}\left(\frac{q_{-}}{2}\right)}}
\nonumber
\end{align}
with $\tau\equiv2t_{\bot}/\Delta$. Note that, in contrast to the normal
state considered in the previous section, the interlayer susceptibility
$\mathcal{X}_{n}^{S}$ is finite for the quadratic in-layer spectrum corresponding
to the energy-independent DoS. It has the negative sign leading to antiferromagnetic
interactions between the magnetic layers. The sum $\sum_{n=-\infty}^{\infty}\mathcal{X}_{n}^{S}=-\nu_{2D}$
meaning that for the total interlayer susceptibility we have $\sum_{n=-\infty}^{\infty}\left(\mathcal{X}_{n}^{S}\!+\!\mathcal{X}_{n}^{N}\right)=0$.
This is the well-known result for vanishing uniform spin susceptibility
in the superconducting state. In the range $\tau\ll1$, we obtain $\mathcal{X}_{n}^{S}/\nu_{2D}=\!-\left(1\!-\frac{1}{6}\tau^{2}\right)\delta_{n}\!-\frac{1}{12}\tau^{2}\delta_{|n|-1}$.
Therefore, in the limit of weak tunneling, $t_{\bot}\ll\Delta$, $\mathcal{X}_{n}^{S}$
is only sizable for the nearest neighbors, $n\!=\!1$, as for the normal part, and $|\mathcal{X}_{1}^{S}|\!\propto\! (t_{\bot}/\Delta)^2\nu_{2D}\! \ll\! \mathcal{X}_{0}$. 
On the other hand, $\mathcal{X}_{n}^{S}$ has a nonmonotonic dependence on $t_{\bot}/\Delta$.
For example, the absolute value of $\mathcal{X}_{1}^{S}$ reaches
maximum at $t_{\bot}\!\approx2.4\Delta$. The plots of $\mathcal{X}_{n}^{S}/\nu_{2D}$
for several values of $t_{\bot}/\Delta$ are shown in Fig.~\ref{Fig:XSCn} (lower left).

The asymptotics at large $n$ 
\begin{equation}
\mathcal{X}_{n}^{S} \approx-\nu_{2D}\frac{\exp\left(-\frac{\Delta}{t_{\bot}}n\right)}{2n},\label{eq:IntSuscSCAsymp}
\end{equation}
is also evaluated in the Appendix \ref{App:DerivSCSusc}. This asymptotics is compared with the exact dependence for $t_{\bot}\!=\!\Delta$ in the upper left plot in Fig.~\ref{Fig:XSCn}.  Note that
the ratio $t_{\bot}/\varDelta$ is approximately equal to the ratio of the $c$-axis
coherence length and the interlayer period. We see that in the case $t_{\bot}\gtrsim\Delta$
superconductivity introduces the long-range interaction between the magnetic
layers. Such behavior is unique for the interlayer RKKY interaction between ferromagnetic layers.

\section{Generalization to multiple-band materials}
\label{sec:multiband}

The Fermi surface of iron pnictides is composed of several holelike
sheets near the Brillouin-zone center and electronlike sheets at the
Brillouin-zone edge. The bands crossing the Fermi level are mostly
composed from the iron d-orbitals. The corresponding Fermi surfaces
are typically open, except near the Lifshitz transitions. In the case
of hole-doped 1144 materials, the band-structure calculations suggest
that six hole bands and four electron bands cross the Fermi level
\cite{LochnerPhysRevB.96.094521,XuCommPhys2019}.

The results of this paper can be straightforwardly generalized to
the multiband case. A band with the index $j$ is characterized by
set of relevant parameters: hopping integral $t_{\bot,j}$, 2D density
of states $\nu_{2D,j}$, superconducting gap $\Delta_{j}$, and exchange
interaction with the local moments $\mathcal{J}_{j,nm}$. The gap
parameters in the electron and hole bands may have opposite signs
($s_{\pm}$ state). This feature has no influence of the phenomena
studied in this paper and $\Delta_{j}$ notate the absolute values
of the gaps. The interlayer RKKY interactions in Eq.~\eqref{eq:EffExchange}
can be obtained by the summation of the band contributions,
\begin{equation}
\mathcal{I}_{m-m^{\prime}}=\sum_{j}\sum_{n,n^{\prime}}\mathcal{X}_{j,n-n^{\prime}}\mathcal{J}_{j,nm}\mathcal{J}_{j,n^{\prime}m^{\prime}}.\label{eq:ExchMultiband}
\end{equation}
The normal and superconducting contributions to the partial interlayer
susceptibilities $\mathcal{X}_{j,n}$ are similar to the corresponding
single-band results in Eqs.~\eqref{eq:NN-SecondDir} and \eqref{eq:IntSuscSCPres}
\[
\mathcal{X}_{j,1}^{N}  \approx\frac{\nu_{2D,j}^{\prime\prime}}{6}t_{\bot,j}^{2},\ \
\mathcal{X}_{j,n}^{S}  =\nu_{2D,j}\mathcal{S}_{n}(2t_{\bot,j}/\Delta_{j}),
\]
where the function $\mathcal{\mathcal{S}}_{n}(\tau)$ is defined in
Eq.~\eqref{eq:IntSuscSCPres}. As these components are controlled
by very different electronic parameters, the dominating contributions
to the normal and superconducting parts of $\mathcal{I}_{m-m^{\prime}}$
may come from different bands. For more complicated interlayer tunneling
mechanisms, the $z$-axis spectrum $\varepsilon_{z,j}$ may have non-cosine
form and depend on the in-plane momentum $\boldsymbol{p}$. In this
case the parameter $t_{\bot,j}$ in the above equations has to be
replaced with $\left\langle \left|\int_{-\pi}^{\pi}\frac{dq}{2\pi}\varepsilon_{z,j}(\boldsymbol{p}_{F,j},q)\exp\left(iq\right)\right|^{2}\right\rangle ^{1/2}$,
where the averaging has to be taken over the in-plane Fermi momentum
$\boldsymbol{p}_{F,j}$.

As the normal contribution to the interlayer susceptibility is ferromagnetic and  short-range and the superconducting contribution is antiferromagnetic and long-range for $t_{\bot,j}>\Delta_{j}$, the interactions between the magnetic layers are frustrated. This is the main reason for the formation of the helical magnetic structures considered in the next section.

\section{Energy of interlayer helical structure}\label{sec:HelixEn}

\begin{figure}
	\includegraphics[width=3.3in]{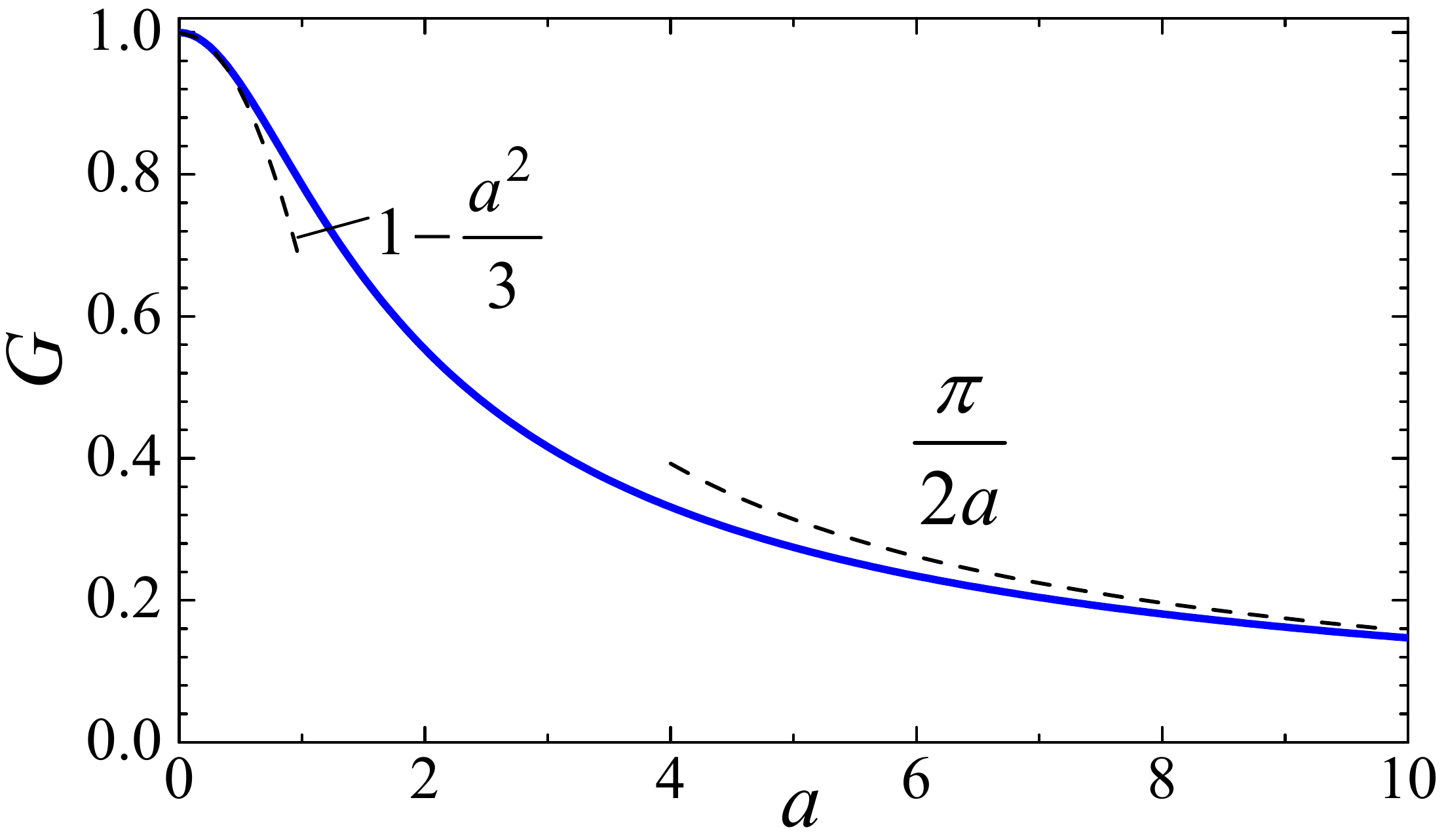}
	\caption{The plot of the function $G(a)$ defined by Eq.\ \eqref{eq:Ga}, which determines
		the dependence of the superconducting interlayer energy on the modulation
		wave vector $Q$, Eq.~\eqref{eq:FMSQRes}.}
	\label{Fig:Ga}
\end{figure}
We consider the magnetic structure in the form of a helix
\[
\left\langle S_{m}^{X}\right\rangle =S_{0}\cos\left(Qm\right),\left\langle S_{m}^{Y}\right\rangle =S_{0}\sin\left(Qm\right),\left\langle S_{m}^{Z}\right\rangle =0.
\]
In this case the interlayer-interaction energy, Eq.~\eqref{eq:FMS},
becomes
\begin{equation}
F_{\mathrm{h}}(Q)=-S_{0}^{2}\sum_{l=1}^{\infty}\mathcal{I}_{l}\cos\left(Ql\right).\label{eq:FMSQ}
\end{equation}

As this work is motivated by the magnetic 1144 iron arsenides, we
now evaluate the energy of helical structure for layer's arrangement
realized in these compounds illustrated in Fig.~\ref{Fig:XSCn}. In this case, the relation between the interlayer exchange interaction $\mathcal{I}_{l}$ and interlayer susceptibility $\mathcal{X}_{n}$ is given by Eq.~\eqref{eq:IlXl1144}. For illustration, we consider here a single-band case. As discussed in the previous section, generalization to multiple bands is straightforward.
In the normal state, we can take into account only the nearest-neighbor
interlayer susceptibility $\mathcal{X}_{1}^{N}$ yielding
\[
\mathcal{I}_{1}^{N}\approx\mathcal{J}^{2}\mathcal{X}_{1}^{N},
\]
where $\mathcal{X}_{1}^{N}$ is given by Eq.~\eqref{eq:NN-SecondDir}.
Therefore, the normal contribution to the energy, Eq.~\eqref{eq:FMSQ}
is 
\begin{equation}
F_{\mathrm{h}}^{N}(Q)\approx-h_{0}^{2}\mathcal{X}_{1}^{N}\cos Q,
\end{equation}
where
\begin{equation}
h_0=\mathcal{J}S_{0}
\end{equation}
is the amplitude of exchange field induced by polarized magnetic ions on the conducting electrons.

The derivation of the superconducting contribution is outlined in
Appendix \ref{App:FMSQ} and leads to the following result
\begin{align}
F_{\mathrm{h}}^{S}(Q)\! & =\frac{\nu_{2D}}{2}h_{0}^{2}\left\{ \left(1\!+\!\cos\frac{Q}{2}\right)G\left[\frac{2t_{\bot}}{\Delta}\sin\frac{Q}{4}\right]\right.\nonumber \\
+ & \left.\left(1\!-\!\cos\frac{Q}{2}\right)G\left[\frac{2t_{\bot}}{\Delta}\cos\frac{Q}{4}\right]\right\} \label{eq:FMSQRes}
\end{align}
with the reduced function
\begin{equation}\label{eq:Ga}
G\left(a\right)=\frac{1}{a}\int_{0}^{\pi}\frac{dq}{\pi}\frac{\ln\left(\sqrt{a^{2}\sin^{2}q+1}+a\sin q\right)}{\sin q\sqrt{a^{2}\sin^{2}q+1}},
\end{equation}
which is plotted in Fig.~\ref{Fig:Ga}. This function has the following asymptotics 
\[
G\left(a\right)\simeq\begin{cases}
1-\frac{1}{3}a^{2}, & \mathrm{for}\:a\ll1\\
\frac{\pi}{2a}, & \mathrm{for}\:a\gg1
\end{cases},
\]
also shown in the plot. Independently on the relation between $t_{\bot}$
and $\Delta$, $F_{\mathrm{h}}^{S}(Q)$ monotonically decreases with increasing
$Q$ indicating that the superconducting contribution by itself favors
the maximum possible modulation vector $Q=\pi$. In two limiting cases,
we obtain 
\begin{widetext}
\begin{equation}
F_{\mathrm{h}}^{S}(Q)\simeq\frac{\nu_{2D}}{2}h_{0}^{2}\times\begin{cases}
2-\frac{1}{3}\frac{4t_{\bot}^{2}}{\Delta^{2}}\sin^{2}\frac{Q}{2}, & \mathrm{for}\:\frac{2t_{\bot}}{\Delta}\ll1\\
\frac{\pi}{2}\frac{\Delta}{t_{\bot}}\left(\cot\frac{Q}{4}\cos\frac{Q}{4}+\tan\frac{Q}{4}\sin\frac{Q}{4}\right), & \mathrm{for}\:\frac{2t_{\bot}}{\Delta}\sin\frac{Q}{4}\gg1
\end{cases}.\label{eq:FMSQAsymp}
\end{equation}
\end{widetext}
In the case $t_{\bot}\!\gg\!\Delta$, the superconducting contribution
increases $\propto1/Q$ with decreasing $Q$ in the range $\Delta/t_{\bot}\ll Q\ll1$,
similar to the isotropic case \citep{BulaevskiiJLTP1980}. %
We proceed with evaluation of the optimal modulation wave vector from the derived energy of the helical state,
$F_{\mathrm{h}}(Q)=F_{\mathrm{h}}^{N}(Q)+F_{\mathrm{h}}^{S}(Q)$.

\section{Optimal modulation wave vector}\label{sec:OptMod}

The helical structures with the modulation wave vector in the range $0\!<\!Q\!<\!\pi$
 may realize only if the interlayer hopping $t_{\bot}$ is either
comparable with or larger than the gap $\Delta$. We limit ourselves
to the analysis of the case $t_{\bot}\!\gg\!\Delta$. In addition to the normal
and superconducting RKKY contributions considered in the previous
sections, the ground-state configuration of the localized moments
is also affected by the in-plane four-fold anisotropy described by the
single-spin energy $-K_{4}\left(S_{x}^{4}+S_{y}^{4}\right)$. For
a simple helical structure this gives the contribution to the energy
per layer and per unit area, $-\mathcal{K}_{4}\left\langle \cos^{4}\left(Qm\right)+\sin^{4}\left(Qm\right)\right\rangle _{m}=-\frac{\mathcal{K}_{4}}{4}\left(3+\left\langle \cos\left(4Qm\right)\right\rangle _{m}\right)$,
where $\mathcal{K}_{4}=K_{4}S_{0}^{4}n_{M}$ and $n_{M}$ is the moment's
density per unit area. Therefore we can write the energy of the helical
structure as
\begin{align}
F_{\mathrm{h}}(Q)=&-\mathcal{A}_{N}\cos Q+\mathcal{A}_{S}\frac{\cos^{3}\!\left(\frac{Q}{4}\right)\!+\!\sin^{3}\!\left(\frac{Q}{4}\right)}{\sin\left(\frac{Q}{4}\right)\cos\left(\frac{Q}{4}\right)}\nonumber\\
-&\frac{\mathcal{K}_{4}}{4}\left\langle \cos\left(4Qm\right)\right\rangle _{m}\label{eq:TotalEnHelix}
\end{align}
with $\mathcal{A}_{N}\!\approx\frac{1}{6}h_{0}^{2}\nu_{2D}^{\prime\prime}t_{\bot}^{2}$
and $\mathcal{A}_{S}\!=\frac{\pi}{4}h_{0}^{2}\nu_{2D}\frac{\Delta}{t_{\bot}}$ following from the previous-section results (for multiple-band case,
$\mathcal{A}_{N}\!\approx\frac{1}{6}S_{0}^{2}\sum_{j}\mathcal{J}_{j}^{2}\nu_{2D,j}^{\prime\prime}t_{\bot,j}^{2}$
and $\mathcal{A}_{S}\!=\frac{\pi}{4}S_{0}^{2}\sum_{j}\mathcal{J}_{j}^{2}\nu_{2D,j}\frac{\Delta_{j}}{t_{\bot,j}}$).
Within this simple model, the anisotropic contribution is only finite
for $Q=\pi/2$ and $\pi$. Without the anisotropy term, the optimal
modulation vector $Q_{o}$ continuously changes as function of the
ratio $\mathcal{A}_{S}/\!\mathcal{A}_{N}$. In the presence of the
four-fold anisotropy, $Q_{o}$ is locked to the values $\pi/2$ and
$\pi$ within finite ranges of $\mathcal{A}_{S}/\!\mathcal{A}_{N}$.
These ranges can be approximately estimated by comparison the energies
of commensurate and incommensurate configurations. 

For incommensurate structures with optimal modulation wave vector
$Q_{o}\neq\pi/2,\pi$, the ground-state energy is 
\begin{equation}
F_{\mathrm{h}}(Q_{o})\!=\!\mathcal{A}_{N}\min_{Q}\left[-\cos Q+\frac{\mathcal{A}_{S}}{\mathcal{A}_{N}}
\frac{\cos^{3}\!\left(\frac{Q}{4}\right)\!+\!\sin^{3}\!\left(\frac{Q}{4}\right)}
{\sin\left(\frac{Q}{4}\right)\cos\left(\frac{Q}{4}\right)}\right]
\label{eq:FQInc}
\end{equation}
where, in a single-band case,
\begin{equation}
\frac{\mathcal{A}_{S}}{\mathcal{A}_{N}}=\frac{3\pi}{2}\frac{\nu_{2D}\Delta}{\nu_{2D}^{\prime\prime}t_{\bot}^{3}}.
\label{eq:ASANratio}
\end{equation}
On the other hand, the energies for $Q_{o}=\pi/2$ and $\pi$ are
\begin{align}
F_{\mathrm{h}}(\pi/2) & =C_{\pi\!/\!2}\mathcal{A}_{S}-\frac{\mathcal{K}_{4}}{4}\label{eq:FQhpi}\\
F_{\mathrm{h}}(\pi) & =\mathcal{A}_{N}+\frac{\mathcal{A}_{S}}{\sqrt{2}}-\frac{\mathcal{K}_{4}}{4}\label{eq:FQpi}
\end{align}
with $C_{\pi\!/\!2}\!=\!(1\!+\!1/\sqrt{2})^{3/2}\!+\!(1\!-\!1/\sqrt{2})^{3/2}\!\approx\!2.39$. 

\begin{figure}
\includegraphics[width=3.4in]{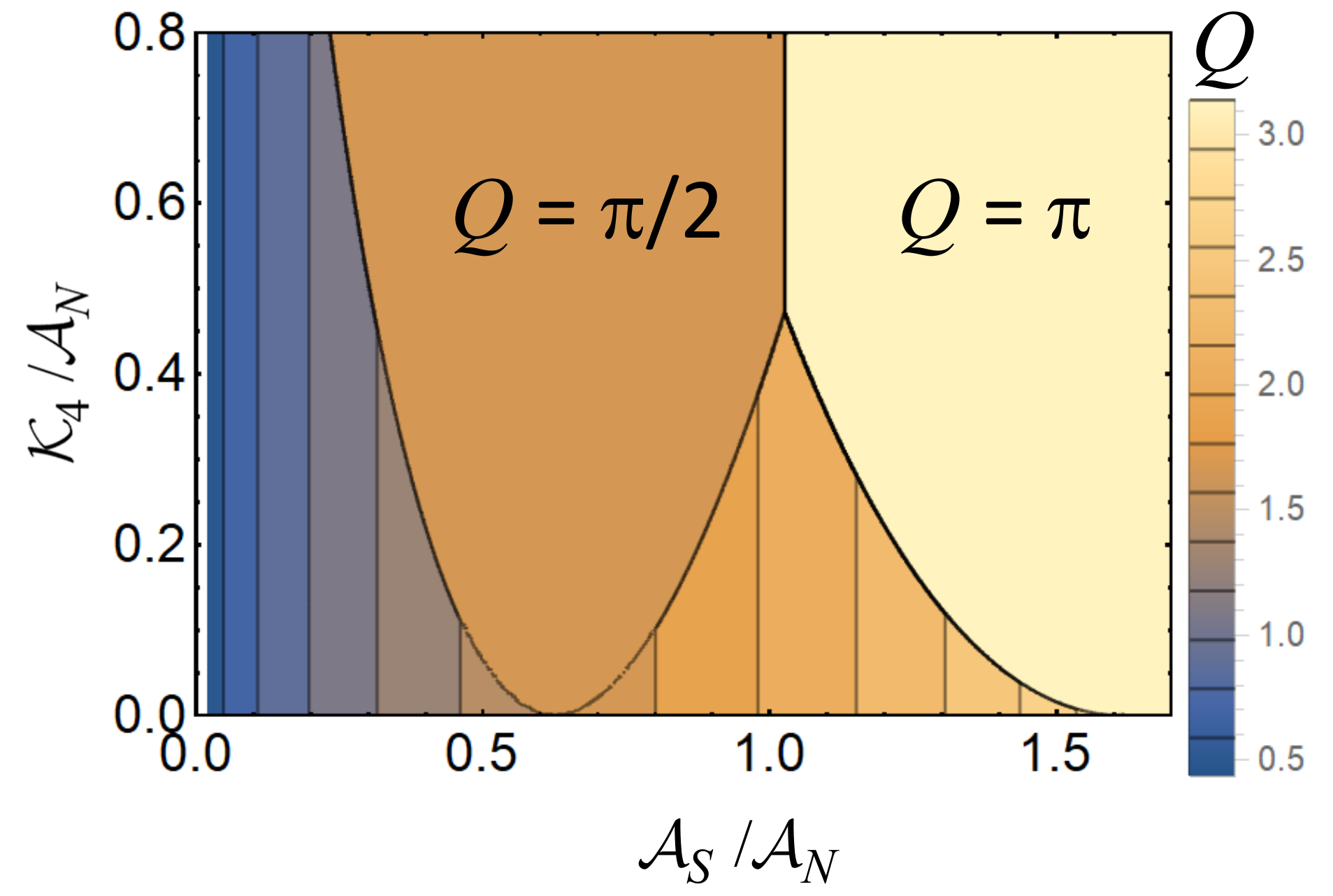}
\caption{Phase diagram for helical structures in layered magnetic superconductors in the plane $\mathcal{A}_{S}/\mathcal{A}_{N}$--$\mathcal{K}_{4}/\mathcal{A}_{N}$,
where the first ratio is given by Eq.~\eqref{eq:ASANratio}. This
diagram is computed for the layer arrangement corresponding to RbEuFe$_{4}$As$_{4}$,
see inset in Fig.~\ref{Fig:XSCn}, and for the case $t_{\bot}\gg\Delta$. }
\label{Fig:HelixPhDiag}
\end{figure}

Figure \ref{Fig:HelixPhDiag} shows the phase diagram following from
comparison of the energies in Eqs.~\eqref{eq:FQInc}, \eqref{eq:FQhpi}
and \eqref{eq:FQpi}. We see that the modulation vector increases
with increasing the ratio $\mathcal{A}_{S}/\!\mathcal{A}_{N}$, and at finite
$\mathcal{K}_{4}$ the commensurate states are realized within finite
ranges of $\mathcal{A}_{S}/\!\mathcal{A}_{N}$. In particular, at
$\mathcal{K}_{4}\!=\!0$ the $\pi/2$ helix is realized at $\mathcal{A}_{S}/\!\mathcal{A}_{N}\approx0.625$
and the range of $\mathcal{A}_{S}/\!\mathcal{A}_{N}$, where this
phase is locked rapidly increases with increasing $\mathcal{K}_{4}$.
Note that a simple energy comparison only gives an approximate location
of the lines, because in the vicinity of transition a helix does not
give the ground state. Consequently, the energy comparison gives somewhat
larger extent of the commensurate states. The actual transition from
commensurate to incommensurate state occurs via formation of the soliton
lattice \citep{BakRoPP1982,BookSolitons}. The detailed investigation
of transitions is beyond the scope of this paper. 

\section{Summary and discussion}
\label{sec:Summary}

In conclusion,  we investigated the interlayer RKKY interactions and equilibrium magnetic structure for a material composed of superconducting and ferromagnetic layers in the situation when the finite coupling between the magnetic layers is mediated by tunneling between the conducting layers.  
We demonstrated that for a shallow band with the weakly-nonparabolic spectrum, the normal-state contribution to the interlayer RKKY interaction is ferromagnetic and short-range. On the other hand, the superconducting part is antiferromagnetic and may be long-range if the hopping integral exceeds the superconducting gap. As a result of the competition between these two contributions, the ground-state magnetic configuration may be a spiral, similar to isotropic case \citep{BulaevskiiJLTP1980}.  
On the phenomenological level, the mechanism of the spiral formation is the same as in the Heisenberg model with frustrating exchange interactions, see, e.g., Refs.~\citep{Nagamiya1968,JohnstonPhysRevB.91.064427}. Our model provides a natural physical mechanism for such a frustration.
In absence of the anisotropy with respect to in-plane rotations of the magnetic moments, the angle between them in the neighboring layers would depend continuously on the model parameters.
The finite four-fold moment-rotation anisotropy may lock this angle to 90$^\circ$, as it was observed in RbEuFe$_{4}$As$_{4}$.

The superconducting transition in RbEuFe$_{4}$As$_{4}$ sinks below the magnetic transition with increasing pressure\cite{JacksonPhysRevB.98.014518,XiangPhysRevB.99.144509}. 
Our model predicts the ferromagnetic alignment of Eu moments in the normal state.  Therefore, establishing the magnetic structure in the normal state at high pressures provides an essential test for the model. 

Our minimum model does not capture a complicated multiple-band structure
of RbEuFe$_{4}$As$_{4}$ obtained by \textit{ab initio} calculations\cite{XuCommPhys2019,IidaPhysRevB.100.014506}.
These calculations suggest that six hole bands and four electron bands
cross the Fermi level. All Fermi surfaces are open, in spite of noticeable
variations the electronic parameters between the bands. %
The straightforward generalization of the model to the case of multiple open Fermi surfaces is discussed in Sec.~\ref{sec:multiband}. Unfortunately, the DFT-based band-structure calculations still have large inaccuracies. For example, the normal-state specific heat coefficient calculated in Ref.~\cite{XuCommPhys2019} is $\sim$ 5-6 times smaller than the experimental value, most likely due to the correlation effects. Do to these inaccuracies and lack of direct experimental information, the relevant parameters can only estimated very approximately: the interlayer hopping $t_{\perp}=10-40$ meV, in-layer band width $W=0.2-1$ eV, gap parameter $\Delta=2-10$ meV, and the exchange-field amplitude  $h_{0}=S_{0}\mathcal{J}=0.5-1$ meV. Due to such large uncertainties, the quantitative analysis does not look feasible yet. Our consideration only illustrates an essential physical mechanism responsible for the formation of the spiral magnetic structure in RbEuFe$_{4}$As$_{4}$ and similar materials and provides the basis for more realistic descriptions.

\begin{acknowledgments}
	I would like to thank U.\ Welp, Z.\ Islam, O.\ Chmaissem, V. K. Vlasko-Vlasov, and S.\ Rosenkranz for discussions of experimental data and physics of the magnetic iron arsenides.  
	This work was supported by the US Department of Energy, Office of Science, Basic Energy Sciences, Materials Sciences and Engineering Division.
\end{acknowledgments}
\appendix
\begin{widetext}

\section{Derivation of interlayer susceptibility in superconducting state\label{App:DerivSCSusc}}

In this appendix we present the derivation details for the susceptibility
integrated over in-plane coordinates, Eqs.~\eqref{eq:IntSuscDef} and \eqref{eq:IntSuscSCDef}. Using the presentation
\[
 \int d\boldsymbol{r}\mathcal{G}^{2}\left(\omega_{s},\boldsymbol{r},n\right)=\int\frac{d^{2}\boldsymbol{p}}{(2\pi)^{2}}\mathcal{G}^{2}\left(\omega_{s},\boldsymbol{p},n\right)
 =\int d\xi\nu_{2D}(\xi)\int_{-\pi}^{\pi}\frac{dq}{2\pi}\int_{-\pi}^{\pi}\frac{dq^{\prime}}{2\pi}\mathcal{G}\left(\omega_{s},\xi,q\right)\mathcal{G}\left(\omega_{s},\xi,-q'\right)\exp\left[i\left(q-q^{\prime}\right)n\right]
\]
with 
\[
\mathcal{G}\left(\omega_{s},\xi,q\right)  =\frac{i\omega_{s}+\xi+\varepsilon_{z}(q)}{\omega_{s}^{2}+\left[\xi+\varepsilon_{z}(q)\right]^{2}+\Delta^{2}},
\]
and similar presentation for the anomalous part $\int d\boldsymbol{r}\mathcal{F}^{2}\left(i\omega_{s},\boldsymbol{r},n\right)$,
we transform $\mathcal{X}_{n}$ in Eq.~\eqref{eq:IntSuscSCDef} to
\[
\mathcal{X}_{n}=-\int d\xi\nu_{2D}(\xi)\int_{-\pi}^{\pi}\frac{dq}{2\pi}\int_{-\pi}^{\pi}\frac{dq^{\prime}}{2\pi}\exp\left[i\left(q-q^{\prime}\right)n\right]\mathcal{R}\left[\xi+\varepsilon_{z}(q),\xi+\varepsilon_{z}(q^{\prime})\right],
\]
where
\[
\mathcal{R}\left(\varsigma,\varsigma^{\prime}\right)=-T\sum_{\omega_{s}}\frac{\omega_{s}^{2}-\varsigma\varsigma^{\prime}-\Delta^{2}}{\left(\omega_{s}^{2}+\varsigma^{2}+\Delta^{2}\right)\left(\omega_{s}^{2}+\varsigma^{\prime2}+\Delta^{2}\right)}.
\]
Performing summation over $\omega_{s}$, we obtain
\begin{align*}
\mathcal{R}\left(\varsigma,\varsigma^{\prime}\right) & =-\frac{1}{2\left(\varsigma^{2}-\varsigma^{\prime2}\right)}\left[\frac{\varsigma^{2}+\varsigma\varsigma^{\prime}+2\Delta^{2}}{\sqrt{\varsigma^{2}+\Delta^{2}}}\tanh\left(\frac{\sqrt{\varsigma^{2}+\Delta^{2}}}{2T}\right)-\frac{\varsigma^{\prime2}+\varsigma\varsigma^{\prime}+2\Delta^{2}}{\sqrt{\varsigma^{\prime2}+\Delta^{2}}}\tanh\left(\frac{\sqrt{\varsigma^{\prime2}+\Delta^{2}}}{2T}\right)\right],
\end{align*}
where we used $T\sum_{\omega_{s}}\frac{1}{\omega_{s}^{2}+z^{2}}=\frac{\tanh\left(z/2T\right)}{2z}$.
In the limit $T\rightarrow0$
\begin{align*}
\mathcal{R}\left(\varsigma,\varsigma^{\prime}\right)= & \!-\frac{1}{2\left(\sqrt{\varsigma^{2}\!+\!\Delta^{2}}\!+\!\sqrt{\varsigma^{\prime2}\!+\!\Delta^{2}}\right)}\left[1\!-\frac{\varsigma\varsigma^{\prime}+\Delta^{2}}{\sqrt{\varsigma^{2}\!+\!\Delta^{2}}\sqrt{\varsigma^{\prime2}\!+\!\Delta^{2}}}\right].
\end{align*}
Therefore, at $T=0$ we obtain
\[
\mathcal{X}_{n}\!=\!\int\!d\xi\nu_{2D}(\xi)\!\int\limits _{-\pi}^{\pi}\frac{dq}{2\pi}\int\limits _{-\pi}^{\pi}\frac{dq^{\prime}}{2\pi}\frac{\exp\left[i\left(q-q^{\prime}\right)n\right]}{2\left(\sqrt{\left(\xi\!+\!\varepsilon_{z}\right)^{2}\!+\!\Delta^{2}}\!+\!\sqrt{\left(\xi\!+\!\varepsilon_{z}^{\prime}\right)^{2}\!+\!\Delta^{2}}\right)}\left[1\!-\frac{\left(\xi+\varepsilon_{z}\right)\left(\xi+\varepsilon_{z}^{\prime}\right)+\Delta^{2}}{\sqrt{\left(\xi\!+\!\varepsilon_{z}\right)^{2}\!+\!\Delta^{2}}\sqrt{\left(\xi\!+\!\varepsilon_{z}^{\prime}\right)^{2}\!+\!\Delta^{2}}}\right]
\]
where we use abbreviations $\varepsilon_{z}=\varepsilon_{z}(q)$ and
$\varepsilon_{z}^{\prime}=\varepsilon_{z}(q^{\prime})$. This result
is consistent with the general presentation for susceptibility derived
in the Abrikosov book \citep{inAbrikosovBook}, Eq.~ (21.44). Subtracting
the normal part for $\Delta=0$, we obtain the superconducting contribution
\begin{align*}
\mathcal{X}_{n}^{S}\! & =\!\int\!d\xi\nu_{2D}(\xi)\!\int\limits _{-\pi}^{\pi}\frac{dq}{2\pi}\int\limits _{-\pi}^{\pi}\frac{dq^{\prime}}{2\pi}\exp\left[i\left(q\!-\!q^{\prime}\right)n\right]\left\{ \frac{1}{2\left(\sqrt{\left(\xi\!+\!\varepsilon_{z}\right)^{2}\!+\!\Delta^{2}}\!+\!\sqrt{\left(\xi\!+\!\varepsilon_{z}^{\prime}\right)^{2}\!+\!\Delta^{2}}\right)}\right.\\
\times & \left.\left[1\!-\frac{\left(\xi+\varepsilon_{z}\right)\left(\xi+\varepsilon_{z}^{\prime}\right)+\Delta^{2}}{\sqrt{\left(\xi\!+\!\varepsilon_{z}\right)^{2}\!+\!\Delta^{2}}\sqrt{\left(\xi\!+\!\varepsilon_{z}^{\prime}\right)^{2}\!+\!\Delta^{2}}}\right]-\frac{1\!-\mathrm{sign}\left(\xi+\varepsilon_{z}\right)\mathrm{sign}\left(\xi+\varepsilon_{z}^{\prime}\right)}{2\left(\left|\xi\!+\!\varepsilon_{z}\right|\!+\!\left|\xi\!+\!\varepsilon_{z}^{\prime}\right|\right)}\right\} .
\end{align*}
We remind that the normal part vanishes for energy-independent DoS
at $n\neq0$. The superconducting contribution, however, remains finite
allowing us to neglect DoS energy dependence in it which yields

\begin{align}
\mathcal{X}_{n}^{S}\! & =\nu_{2D}\!\int\limits _{-\pi}^{\pi}\frac{dq}{2\pi}\int\limits _{-\pi}^{\pi}\frac{dq^{\prime}}{2\pi}\!\int d\xi\exp\left[i\left(q\!-\!q^{\prime}\right)n\right]\left\{ \frac{1}{2\left(\sqrt{\left(\xi\!+\frac{\varepsilon_{z-}}{2}\right)^{2}\!+\Delta^{2}}\!+\sqrt{\left(\xi\!-\frac{\varepsilon_{z-}}{2}\right)^{2}\!+\Delta^{2}}\right)}\right.\nonumber \\
 & \times\left.\left[1\!-\!\frac{\xi^{2}-\varepsilon_{z-}^{2}/4+\Delta^{2}}{\sqrt{\left(\xi\!+\frac{\varepsilon_{z-}}{2}\right)^{2}\!+\Delta^{2}}\sqrt{\left(\xi\!-\frac{\varepsilon_{z-}}{2}\right)^{2}\!+\Delta^{2}}}\right]-\frac{\Theta\left(\frac{\left|\varepsilon_{z-}\right|}{2}-\xi\right)}{\left|\varepsilon_{z-}\right|}\right\} .\label{eq:AppInSuscSC}
\end{align}
with $\varepsilon_{z-}\!=\!\varepsilon_{z}\!-\!\varepsilon_{z}^{\prime}$. 
Making substitution $\xi=\frac{\varepsilon_{z-}}{2}x$, we present
the dimensionless ratio $\mathcal{X}_{n}^{S}/\nu_{2D}$ as
\begin{align*}
 & \mathcal{X}_{n}^{S}/\nu_{2D}=\int_{-\pi}^{\pi}\frac{dq}{2\pi}\int_{-\pi}^{\pi}\frac{dq^{\prime}}{2\pi}\exp\left[i\left(q-q^{\prime}\right)n\right]G[2\Delta/|\varepsilon_{z-}|],\\
 & G[a]=\frac{1}{2}\int\limits _{-\infty}^{\infty}dx\left\{ \left[1-\frac{x^{2}-1+a^{2}}{\sqrt{\left(x+1\right)^{2}+a^{2}}\sqrt{\left(x-1\right)^{2}+a^{2}}}\right]\frac{1}{\sqrt{\left(x+1\right)^{2}+a^{2}}+\sqrt{\left(x-1\right)^{2}+a^{2}}}-\Theta(1-|x|)\right\} \\
 & =-\frac{a^{2}}{\sqrt{a^{2}+1}}\ln\left(\frac{\sqrt{a^{2}+1}+1}{a}\right).
\end{align*}
This gives a useful presentation
\begin{equation}
\frac{\mathcal{X}_{n}^{S}}{\nu_{2D}}=-\int_{-\pi}^{\pi}\frac{dq}{2\pi}\int_{-\pi}^{\pi}\frac{dq^{\prime}}{2\pi}\exp\left[i\left(q-q^{\prime}\right)n\right]\frac{2\Delta^{2}/\left|\varepsilon_{z-}\right|}{\sqrt{\Delta^{2}+\varepsilon_{z-}^{2}/4}}\ln\left(\frac{\sqrt{\Delta^{2}+\varepsilon_{z-}^{2}/4}+\left|\varepsilon_{z-}\right|/2}{\Delta}\right).\label{eq:AppIntegrSuscInterm}
\end{equation}
For simple tunneling spectrum $\varepsilon_{z-}=4t_{\bot}\sin\left(q_{+}\right)\sin\left(\frac{q_{-}}{2}\right)$
with $q_{+}=\frac{q+q^{\prime}}{2}$ and $q_{-}=q-q^{\prime}$. This
allows us to transform Eq.~\eqref{eq:AppIntegrSuscInterm} to the
following form
\begin{align}
\frac{\mathcal{X}_{n}^{S}}{\nu_{2D}} & =-\int_{0}^{\pi}\frac{dq_{+}}{\pi}\int_{0}^{\pi}\frac{dq_{-}}{\pi}\frac{\cos\left(q_{-}n\right)\ln\left(\sqrt{1+\tau^{2}\sin^{2}q_{+}\sin^{2}\left(\frac{q_{-}}{2}\right)}+\tau\sin q_{+}\sin\left(\frac{q_{-}}{2}\right)\right)}{\tau\sin q_{+}\sin\left(\frac{q_{-}}{2}\right)\sqrt{1+\tau^{2}\sin^{2}q_{+}\sin^{2}\left(\frac{q_{-}}{2}\right)}}\label{eq:AppIntSuscSCPres}
\end{align}
with $\tau\equiv2t_{\bot}/\Delta$. This result is equivalent to Eq.~\eqref{eq:IntSuscSCPres} of the main text. In particular, for $\tau\ll1$,
using expansion $\ln\left(\sqrt{1+x^{2}}+x\right)\approx x-\frac{1}{6}x^{3}$,
we obtain
\begin{align}
\frac{\mathcal{X}_{n}}{\nu_{2D}} & =-\left(1-\frac{1}{6}\tau^{2}\right)\delta_{n}-\frac{1}{12}\tau^{2}\delta_{|n|-1}.
\end{align}
Therefore, in the case $t_\bot\!\ll \!\Delta$ the superconducting contribution to the RKKY interaction is short-range, similar to the normal part.

The asymptotics at large $n$ is determined by the region $q_{-}\ll1$.
\begin{align*}
\frac{\mathcal{X}_{n}^{S}}{\nu_{2D}} & \approx-\int_{0}^{\pi}\frac{dq_{+}}{\pi}\int_{0}^{\infty}\frac{dq_{-}}{\pi}\cos\left(q_{-}n\right)\frac{\ln\left(\sqrt{1+\tau^{2}\sin^{2}q_{+}\left(\frac{q_{-}}{2}\right)^{2}}+\tau\sin q_{+}\left(\frac{q_{-}}{2}\right)\right)}{\tau\sin q_{+}\left(\frac{q_{-}}{2}\right)\sqrt{1+\tau^{2}\sin^{2}q_{+}\left(\frac{q_{-}}{2}\right)^{2}}}.
\end{align*}
Making substitution $k\!=\!\tau\sin\left(q_{+}\right)\frac{q_{-}}{2}$,
we obtain presentation
\begin{align*}
\frac{\mathcal{X}_{n}^{S}}{\nu_{2D}} & \approx-\frac{2}{\tau\pi}\int_{0}^{\pi}\frac{dq_{+}}{\pi\sin q_{+}}\mathcal{G}\left(\frac{2}{\tau}\frac{n}{\sin q_{+}}\right),\\
\mathcal{G}\left(a\right)= & \int_{0}^{\infty}dk\cos\left(ak\right)\frac{\ln\left(\sqrt{1+k^{2}}+k\right)}{k\sqrt{1+k^{2}}}.
\end{align*}
Deforming the integration contour into the complex plane, we obtain
\begin{align*}
\mathcal{G}\left(a\right) & =\frac{\pi}{2}\int_{1}^{\infty}dz\frac{\exp\left(-az\right)}{z\sqrt{z^{2}-1}}\approx\frac{\pi}{2\sqrt{2}}\exp\left(-a\right)\int_{0}^{\infty}dx\exp\left(-ax\right)\frac{1}{\sqrt{x}}=\frac{\pi^{3/2}}{2\sqrt{2a}}\exp\left(-a\right).
\end{align*}
Therefore
\begin{align*}
\frac{\mathcal{X}_{n}^{S}}{\nu_{2D}} & \approx-\frac{1}{2\sqrt{\pi\tau n}}\int_{0}^{\pi}\frac{dq_{+}}{\sqrt{\sin q_{+}}}\exp\left(-\frac{2}{\tau}\frac{n}{\sin q_{+}}\right)\\
\approx & -\frac{\exp\left(-\frac{2}{\tau}n\right)}{2\sqrt{\pi\tau n}}\int_{-\infty}^{\infty}dx\exp\left(-\frac{n}{\tau}x^{2}\right)=-\frac{\exp\left(-\frac{2}{\tau}n\right)}{2n}
\end{align*}
giving the result in Eq.~\eqref{eq:IntSuscSCAsymp}.

\section{Derivation of superconducting energy of helical structure\label{App:FMSQ}}

The superconducting contribution to the helical-structure energy,
Eq.~\eqref{eq:FMSQ} is
\begin{align}
F_{\mathrm{h}}^{S}(Q) & =-S_{0}^{2}\mathcal{J}^{2}\sum_{l=1}^{\infty}\sum_{\delta,\delta^{\prime}=0,1}\mathcal{X}_{2l-\delta+\delta^{\prime}}^{S}\cos\left(Ql\right),\label{eq:FMSQS}
\end{align}
where $\mathcal{X}_{n}^{S}$ is given by Eq.~\eqref{eq:IntSuscSCPres}.
For further simplification, we use the identity 
\begin{align*}
\sum_{\delta,\delta^{\prime}=0,1}\!\sum_{l=1}^{\infty}\cos\left[q_{-}\left(2l\!-\!\delta\!+\!\delta^{\prime}\right)\right]\cos\left(Ql\right) & \!=\frac{\pi}{2}\left(1\!+\!\cos q_{-}\right)\left\{ \sum_{m=-\infty}^{\infty}\!\left[\delta(q_{-}\!+\frac{Q}{2}-\!\pi m)+\delta(q_{-}\!-\frac{Q}{2}-\!\pi m)\right]\!-\!2\right\} .
\end{align*}
As $0<q_{-},Q<\pi$, only nonzero $\delta$-function terms are with
$q_{-}\!=Q/2$ and $q_{-}\!=\pi-Q/2$, which gives the $Q$-dependent
part of energy as 
\begin{align}
F_{\mathrm{h}}^{S}(Q) & =\frac{1}{2}\nu_{2D}S_{0}^{2}\mathcal{J}^{2}\int_{0}^{\pi}\frac{dq_{+}}{\pi}\left[\left(1+\cos\frac{Q}{2}\right)
\frac{\ln\left(\sqrt{1\!+\!\tau^{2}\sin^{2}q_{+}\sin^{2}\!\left(\frac{Q}{4}\right)}\!+\!\tau\sin q_{+}\sin\!\left(\frac{Q}{4}\right)\right)}
{\tau\sin q_{+}\sin\left(\frac{Q}{4}\right)\sqrt{1+\tau^{2}\sin^{2}q_{+}\sin^{2}\left(\frac{Q}{4}\right)}}\right.\nonumber\\
+ & \left.\left(1-\cos\frac{Q}{2}\right)\frac{\ln\left(\sqrt{1\!+\!\tau^{2}\sin^{2}q_{+}\cos^{2}\!\left(\frac{Q}{4}\right)}\!+\!\tau\sin q_{+}\cos\!\left(\frac{Q}{4}\right)\right)}{\tau\sin q_{+}\cos\left(\frac{Q}{4}\right)\sqrt{1+\tau^{2}\sin^{2}q_{+}\cos^{2}\left(\frac{Q}{4}\right)}}\right].
\end{align}
This result is equivalent to Eq.~\eqref{eq:FMSQRes} in the main text.
\end{widetext}

\bibliography{FM-SC-Helic-Lay}

\end{document}